\documentclass[lnbip]{svmultln}
\usepackage{makeidx}  
\begin{document}
\mainmatter              
\title{Physical time and thermal clocks}
\titlerunning{Physical time and thermal clocks}  
%
\author{Claudio Borghi}
\authorrunning{Claudio Borghi}   
%
\tocauthor{Claudio Borghi}
\institute{Liceo Scientifico Belfiore, via Tione 2, Mantova, Italy}

\maketitle              

\begin{abstract}        
In this paper I discuss the concept of time in physics. I consider the thermal time hypothesis and I claim that thermal clocks and atomic clocks measure different physical times, whereby thermal time and relativistic time are not compatible with each other. This hypothesis opens the possibility of a new foundation of the theory of physical time, and new perspectives in theoretical and philosophical researches.   
\end{abstract}
\vskip 1cm
\section{Introduction}

The matter can be summarized in the following points:
\begin{itemize}
\item considerations about the concept of physical time in the theoretical frameworks of classical mechanics, Einsteinian relativity, quantum mechanics and quantum gravity;
\item reflections about the peculiar nature of time in thermodynamics in the light of the thermal time hypothesis;
\item formulation of the hypothesis about the different behaviour of thermal clocks with respect to relativistic clocks, and about the consequent nonequivalence between thermal time and relativistic time;
\item reflections about the need of a refoundation of the concept of time in physics, in the light of the behaviour of real clocks employed for measuring durations.
\end{itemize}
The concluding remarks intend to develop the consequences of the probable existence of different physical times, which is configured as potentially revolutionary. 


\vskip 1cm
\section{Classical, relativistic and quantum time}
In the framework of Einsteinian relativity it must be distinguished the coordinate time \textit {t}, which appears, for example, as argument of the variable field $g_{\mu\nu}(x,t)$, from the proper time $\tau$ measured by a clock along a given world line. While in Newton's theory real clocks provide a relative and approximate measure\footnote{The measure of durations in Newton is relative and approximate as obtained through devices that simulate the flow of absolute time, by definition mathematic, then ideal.} of the absolute duration, therefore of the variable \textit {t}, in Einstein's theory clocks measure the length $\tau$ of the world line, hence not \textit {t}, which appears to be a simple mathematical label without physical meaning. The lack of an absolute temporal reference means that in relativity the evolution of bodies and phenomena is not a function of an independent and preferential variable as, instead, it happens with Newtonian time, that plays the role of independent parameter to which every evolution is referred. According to Rovelli \cite {rov1}, relativity describes the evolution of observable quantities relative to one another, without conceiving one of them as independent: for instance, given two clocks, one on the ground and one on a satellite, on which are respectively read the pairs of values ($\tau_1$; $\tau_2$), ($\tau'_1$; $\tau'_2$),($\tau''_1$;$\tau''_2$), etc., corresponding to the proper times that each clock measures along its world line, Einstein's theory predicts the values of $\tau_1$ to be associated with the corresponding values of $\tau_2$, without the need of considering any of them as independent time variable. Although Einstein's theory of proper time implies an overcoming of the Newtonian concept of absolute time, in the light of the inexistence of a same time reference shared by all the observers, it leaves open the problem of the physical meaning of the measurement of durations provided by real clocks. At this regard, Brown \cite {bro} emphasizes that relativistic clocks do not measure time as, for example, the thermometers measure the temperature or the ammeters measure the electric current: their behaviour correlates with some aspects of spacetime, but not in the sense that spacetime acts on them in the way a heat bath acts on a thermometer, or the way a quantum system acts on a measuring device. In fact, the property of relativistic clocks of recording a duration between two fixed extreme events as measure of the length of a world line clearly does not mean that relativity exhausts inside it every possible operational definition of time. The systematic and experimental doubts raised by Brown about the possibility that relativistic clocks do not measure in the classical sense, since they merely record the correlation between the instrument and particular aspects of the spacetime structure, implicitly open the door to the recognition of a potential multiple reality of time, that requires new conceptual and operational tools to be explored. As concerns quantum theories it is remarkable that, according to Barbour \cite {bar}, in quantum mechanics the physical reality can be interpreted as a set of snapshots without actual evolution, while, according to Rovelli \cite {rov2}, quantum gravity reduces the fundamental reality of phenomena to a network of relations between quantum covariant fields, in which the reality of time seems to be illusory. In dialectic relationship with Rovelli's need to forget time in the theoretical framework of relativity, quantum mechanics and quantum gravity, in the following sections we will explore the meaning and the implications of the irreversible reality of time that autonomously emerges from thermodynamics. 
\vskip 1cm
\section{Time and irreversibility}
On closer inspection, the fact that in all the fundamental theories the concept of irreversibility of the evolution, that real clocks record when their measures are linked to irreversible phenomena, is neglected, does not mean that time can be forgotten, but only that these theories do not refer to an operational concept of time linked to irreversible transformations. In this optics it is fundamental the distinction between the internal evolution of bodies and the evolution linked to the variation of their relative position as it is described in mechanics: the true operational essence of physical time must in fact be searched in thermodynamics, where it is generated inside the measuring instruments as a necessary product of an irreversible evolution. The description of physical phenomena provided by thermodynamics seems implicitly to reveal the existence of a dimension of time different from the one that emerges from the fundamental theories: the following hypothesis is the possible foundation of a new conceptual exploration of real phenomena and of a new operational definition of time.
\vskip 1cm
\section{Thermal time hypothesis }   
Introducing the concept of recovery of time, Rovelli \cite {rov1} suggests that the familiar aspects of time, related to the perception of flow, to the impossibility of going against the evolutionary tide, etc., are not of mechanical but of thermodynamical nature. In a certain sense they emerge at a theoretical level where we statistically describe a physical system with a large number of degrees of freedom. Rovelli states that in statistical mechanics it is possible to introduce the thermal time hypothesis, according to which, though in nature there is not a preferential time variable \textit {t} and a state of preferential equilibrium a priori identifiable does not exist, since all the variables are on equal footing, if a system is in a given state $\rho$ it is statistically possible to identify a variable $t_\rho$ named thermal time, that can be defined as the preferred variable singled out by the state of the system. The thermal time variable $\rho$ is the parameter of the flow of the quantity $H_\rho$ (thermal Hamiltonian) defined by the equation $H_\rho = -ln\rho$, and we call thermal clock any device whose reading linearly increases as a function of $t_\rho$. This means that thermal time is determined by the statistical state of the system, not by an hypothetical flow that drives it to a preferred statistical state. Calling time a certain variable, therefore, we are not making a statement concerning the fundamental structure of reality, but a statement about the statistical distribution used to describe the macroscopic properties of the system under observation. What we empirically call time is the thermal time of the statistical state in which a system is observed, when it is described as a function of the macroscopic parameters we have chosen. Time is thus the expression of our ignorance of the microstates, a conceptual simplification arising from a high number of variables that chaotically change. This chaotic activity involves an irreversible increasing of the molecular disorder and this connotation of irreversibility differentiates the level of reality described by thermodynamics with respect to that described by quantum mechanics or relativity, in whose frameworks, as previously remarked, time does not have a physical meaning linked to the internal evolution of a system, whether it is a set of interacting fields or particles or a single particle that describes a world line. In thermodynamics comes to light a new operational definition, therefore a new reality of time as linked to the becoming intrinsic to bodies and systems, on which the Newtonian theory of mechanics and gravitation, the Einsteinian theory of relativity and quantum mechanics have never focused the attention.  
\vskip 1cm
\section{Hypothesis about the nonequivalence between thermal clocks and atomic clocks}    
According to Martinetti \cite {mar}, the question of time can be summarized in the need to explain the emergence of time: in quantum mechanics as an abstract flow in the space of observables of the system, in relativity as a geometrical flow of proper time in the four dimensional spacetime. Originally, Connes and Rovelli \cite {con} introduced the concept of thermal time with the intention to answer the question of time in quantum gravity: the idea is to extract, from an equilibrium state, the time “as abstract flow” of quantum mechanics (namely, a flow of automorphisms), then to turn it into the locally unique time “as geometrical flow” of relativity. According to the authors, the way of reconciliation of these radically different flows is offered by thermodynamics. We do not believe that the question consists in the possibility of a reconciliation between two concepts of time that clearly do not refer to a flow of something. We believe that the error is to conceive the physical time as a theoretical concept that refers to something external to clocks, while it is necessary to consider it as a physical quantity generated inside clocks, measurable through the count of the number of reversible periodical phenomena (as it happens in atomic clocks) or through the quantification of an amount of irreversible transformation (as it happens in thermal clocks). Atomic and thermal clocks are in fact radically different in relation to the specific kind of internal transformation: if an atomic clock after a trip returns at the point of departure, the measured duration is not linked to an amount of transformation that has influenced its internal state, that returns identical to the initial one, while in a thermal clock the irreversible phenomenon that allows its operation prevents the spontaneous restoration of its initial state. Since a different operational definition implicitly refers to a different theoretical concept, it needs to differentiate the definition of relativistic time from the definition of thermal time, that could not be compatible with each other. At this regard we explicitly interpret the thermal time hypothesis as the potential discovery of the existence of a reality of time operationally different from that implicated by the other theories, in particular by the relativistic theory of proper time. In the quoted paper Connes and Rovelli remark a suggestive fact: in a special relativistic system, a thermal state breaks the Lorentz invariance. As particular example they refer to the average momentum of a gas that, at finite temperature, defines a preferred Lorentz frame. This means that a thermal bath is at rest only in a particular Lorentz frame, whereby, if we apply to such a state the thermal time hypothesis, we single out a preferred time, namely the Lorentz time of the Lorentz frame in which the thermal bath is at rest. This leads to corroborate the hypothesis that a thermal clock cannot provide measurements in agreement with the relativistic
predictions about the behaviour of real clocks, that have been verified in particular through atomic clocks\footnote{As remarkable we point out the experiments performed by Hafele and Keating \cite {haf}, through four cesium clocks flying on commercial airlines around the Earth (two to the east and two to the west), and by Alley \cite {all}, through three rubidium clocks in flight along a closed path.}.
\vskip 1cm
\section{Conclusions: forget or refound time?}    
The question of time can be reformulated in terms of the probable disagreement between the experimental measurements obtained by relativistic clocks and those provided by thermal clocks in the same experimental situations. We argue that the evolutionary nature of time \cite {bor} emerges at a level of description of real phenomena in  which the model of spacetime is not fundamental: the evolutionary time is measurable as generated by clocks that register durations in agreement with the thermal time hypothesis. It can be deduced that the models of physical reality do not necessarily imply one another, each one providing an autonomous theoretical description based on a logical-operative structure different from the others. It must be emphasized that the operational reality of irreversible time revolutionizes the theoretical framework, as if some phenomenal peculiarities that appear at the macroscopic scale were not explainable by the fundamental theories. In essence, physical time shows a different nature and different properties depending on the level of observation of phenomena, and the discovery of the autonomous reality of thermal time with respect to relativistic time teaches that the physical theories do not always have to yield to reductionist impulses. Though the evolution of physics has been marked by the understanding of empirical phenomena in the light of unitary theoretical principles (for example, electric and magnetic phenomena have found a complete synthesis in Maxwell's theory), nature seems to shy away, in matter of time, the will of reducing the complexity of reality to a unified theoretical framework. In matter of time, therefore, the world can be investigated according to several interpretations, and it does not seem possible to merge them into a ultimate theory: the recent synthesis of quantum gravity, though it is founded on a model in which the contradictions between quantum mechanics and general relativity are, in some respects, brilliantly overcome, does not solve in fact the problem of time, of which it merely contemplates, at a fundamental level, the nonexistence, in the illusion that such reductionist vision can explain its ultimate nature. We believe that the problem does not consist in the possibility of explaining the emergence of irreversibility from a reversible timelessness, but in recognizing as autonomous and not always in communication between them the different levels of knowledge of physical phenomena, and consequently the theoretical models used to describe them. If, therefore, relativity requires to be tested through atomic or light clocks, that allow to probe the structure of spacetime on which its conceptual building has been erected, the devices of thermodynamical nature, as thermal clocks, probe that the emerged reality of time is of evolutionary nature, whereby the measure of a duration is linked to an irreversible evolution that occurs inside the instrument. Irreversibility is the peculiar characteristic of physical reality at the macroscopic scale, where our senses perceive the phenomena immersed in a unidirectional time flow\footnote {To paraphrase the title of a recent Rovelli's essay \cite {rov3}, we can say that the irreversible reality of time appears in thermodynamical form. It must be emphasized that one of the objectives of theoretical and experimental research should be the description and understanding of reality as it appears, recognizing, in every abstract theorisation, a potential reductionist risk, wherever we want to explain the irreversibility of physical phenomena through the reversible phenomena that seem to constitute their elementary structure.}. Every theoretical idealization, from Newton to Einstein as far as to quantum mechanics and quantum gravity, albeit using different mathematical models as instruments to interpret the complexity of phenomena, is founded on a clear indisputable preconception about a reality of time as an abstract flow external to bodies and clocks. Thermodynamics is the only theory in which the evolutionary nature of time is fundamental, still little explored or reduced to a coarse vision in which complex phenomena are resolved in statistical syntheses from which, according to Rovelli, comes to light the ignorance of the microstates and of the fundamental structures upon which (in accordance with quantum mechanics, general relativity and quantum gravity) the physical reality is built, since it postulates the existence of an internal time, of which the other theories have not been able to grasp the novelty and the revolutionary depth. Only through an accurate investigation about the behaviour of real clocks, that measure durations through irreversible phenomena that occur inside them, a refoundation of physical thought about time can be born, from which the inability will probably emerge to describe reality in the light of a unitary conceptual structure.

\vskip 1cm
\textbf{Acknowledgments} A special thanks to Silvio Bergia (University of Bologna) for his precious and concrete attention to the development of these ideas. Thanks also to Carlo Rovelli and Julian Barbour for having discussed some aspects of this theoretical analysis of the question of time. 

\vskip 2cm
%

%

\begin{thebibliography}{10}
\bibitem{rov1}
Rovelli, C.: Forget time, Found. Phys. \textbf {41} (2011)
\bibitem{bro}
Brown, H.: The behaviour of rods and clocks in general relativity, and the meaning of the metric field,  arXiv:0911.4440v1 (2009)
\bibitem{bar}
Barbour, J.: The nature of time,  arXiv:0903.3489 (2008)
\bibitem{rov2} 
Rovelli, C.: Quantum Gravity. Cambridge University Press, Cambridge (2004) 
\bibitem{mar}
Martinetti, P.: Emergence of time in quantum gravity: is time necessarily flowing?, Kronoscope \textbf {13}, 67-84 (2013)
\bibitem{con}
Connes, A., Rovelli, C.: Von Neumann Algebra Automorphisms and Time-Thermodynamics Relation in Generally Covariant Quantum Theories, Class. Quantum Grav. \textbf {11}, 2899-2917 (1994) 
\bibitem{haf}
Hafele, J. C., Keating, R. E.: Around-the-world atomic clocks: predicted relativistic time gains,  Science \textbf {177}, 166-168 (1972)
\bibitem{all}
Alley, C. O.: Relativity and clocks, Proc. XXXIII Ann. Symp. on Frequency Control pp. 4-39, doi.org/10.1109/FREQ.1979.200296 (1979)  
\bibitem{bor}
Borghi, C.: Hypothesis about the nature of time and rate of clocks, Annales de la Fondation Louis de Broglie \textbf{38}, 167 (2013) 
\bibitem{rov3} 
Rovelli, C.: Reality is not as it appears, Raffaello Cortina Editore (2014)  
\end{thebibliography}
\end{document}